%% file: main.tex
\pdfoutput=1
\documentclass[11pt,a4paper]{article}

\usepackage[T1]{fontenc}
\usepackage[utf8]{inputenc}
\usepackage{lmodern}
\usepackage[margin=2.4cm]{geometry}
\usepackage{graphicx}
\usepackage{booktabs}
\usepackage{multirow}
\usepackage{amsmath}
\usepackage{microtype}
\usepackage[hidelinks]{hyperref}
\usepackage{caption}
\input{generated/numbers}

\newcommand{\fitwidth}[1]{\resizebox{\ifdim\width>\textwidth\textwidth\else\width\fi}{!}{#1}}
\newcommand{\sig}[1]{\texttt{#1}}

\title{\bfseries Complexity concentration as a god-class and god-method
detector:\\ a pre-registered test of a scale-free outlier rule against tool
definitions and human labels}

\newcommand{\thedoi}{10.5281/zenodo.23581333}
\newcommand{\doiline}{\ifdefined\thedoi\\\small\texttt{doi:\thedoi}\fi}

\author{Daniel Sagenschneider\\\small OfficeFloor\\\small\texttt{daniel@officefloor.net}\doiline}

\date{October 2026}

\begin{document}
\maketitle

\begin{abstract}
\noindent
God-class and god-method detectors mark concentrated complexity with a
raw-scale threshold, a constant fitted on a corpus and re-fitted when the
population moves.
This paper pre-registers a test of a rule with no raw-scale constant: a unit is
a god unit when its complexity is a robust outlier for its own codebase, a
modified z-score of 3.5 or more. On \nUsable{} repositories in ten languages,
\nJava{} of them Java, and the \mlcqUsable{} Java repositories of MLCQ, the
rule is scored against PMD's GodClass, MLCQ's reviewers and two further
definitions, with signal and cutoff chosen on half the repositories and read on
the rest. LLM-written code may gather its complexity differently, the case for a
detector that needs no re-fitting; the first question is whether one can
match the scale of existing human-written code. The verdict is
inconclusive on both populations, by different arms. The rule ranks PMD's god
classes from the per-file maximum of cognitive complexity at per-repository
AUCs of \primAucMacro{} and \mlcqPmdAucMacro{}. On the main run it is, at its default, non-inferior
within the margin to a cutoff tuned on the same population on recovery, and
flags about twice as many non-god classes; on MLCQ it flags about as many and
falls short of the margin on pooled recovery. Pooled, the raw value ranks god
classes better than the per-repository score, and a cutoff carried between
the populations lost
almost nothing: both are human-written code on one scale, so the test is of
matching that scale, not of surviving a shift. The rule's extra flags are
small classes, just over half holding a method at PMD's own method-level
threshold; its misses are large classes whose complexity is spread over many
methods. Two readers of the source did not agree on what the extra flags are.
Concentration finds where the complexity is; whether that place is a god unit
is a question about coupling, which a registered follow-on study takes up.
\end{abstract}

\section{Introduction}

A god class is a class that does too much. Riel named the heuristic
\cite{riel1996}, Fowler catalogued the smell as Large Class and its
method-level twin as Long Method \cite{fowler1999}, and Lanza and Marinescu
turned the judgement into a detection strategy, a conjunction of metric
thresholds \cite{lanza2006, marinescu2004}. Every threshold-based detector in use
follows that template, and the thresholds are constants on a raw scale: a weighted method
count of 47, a cyclomatic complexity of 14, a cognitive complexity of 15,
fifty lines. The constants are frozen percentiles of a corpus
\cite{lanza2006, alves2010, ferreira2012, oliveira2014}, and a constant
derived from a corpus is a statement about that corpus. It has to be
re-derived when the population moves, and there is evidence that it may be
moving: projects that adopted an LLM coding agent showed a persistent rise in
complexity \cite{he2026cursor}, and on a fixed assignment LLM-generated
projects gathered their complexity into fewer classes, which the
threshold-based detectors missed \cite{zhang2026llm}. A detector whose
constant is fitted to the codebase in front of it would then be worth having,
but agent-written code is not yet available at the scale a pre-registered
test needs, so the first step, and this paper's question, is whether such a
detector can match the single scale of existing human-written code at all.

The alternative is a rule whose constant lives on a standardised axis.
Standardise every unit by its own repository's median and median absolute
deviation and flag the units whose modified z-score is 3.5 or more, Iglewicz
and Hoaglin's rule \cite{iglewicz1993}. The 3.5 is still a constant, but the scale it
is applied to, the median and the MAD, is fitted on the codebase at the time
of use and carries no corpus: the rule is scale-free, not constant-free. The question arose in a companion study
\cite{conserved}, which found the amount of complexity an AI agent produced on
a fixed task steady across architectures and instructions while its placement
moved by factors of three to ten. Placement, not amount, was what a design
decided, and that turned attention from the level of a unit's complexity to
its \emph{concentration}. The claim this paper tests is that a scale-free
relative outlier in a per-unit complexity measure that is not size recovers
the classes and methods that tuned, absolute-threshold definitions mark, and
that human reviewers mark, without a raw-scale constant.

The claim was pre-registered: the primary estimand, the decision rule and its
margins, the families a signal may be chosen from, the sample seed and the
bootstrap were committed and tagged before the confirmatory sample was drawn.
The design is tilted against the claim: every tool oracle is itself an
absolute threshold, and the competitor is a cutoff tuned on the explore half
of the same population, which is the right competitor for the first step,
since matching a tuned constant on its own population is the bar a scale-free
rule must clear before a shifted population is worth testing. The paper has
three findings.

\begin{enumerate}
\item \textbf{The verdict is inconclusive on both populations, by different
  arms.} Against PMD's GodClass on \nJava{} Java repositories and again on
  MLCQ's \mlcqUsable{}, the per-file maximum of cognitive complexity ranks god
  classes at per-repository AUCs of \primAucMacro{} and \mlcqPmdAucMacro{}. On the main run the rule
  at its default is non-inferior to a cutoff tuned on the same population on
  both MCC arms, flags about twice as many non-god classes, and the
  false-positive arm does not resolve; on MLCQ it flags about as many as the
  cutoff but falls short of the margin on pooled MCC. Every point estimate is
  on the cutoff's side, tuning the rule's own constant changes neither
  verdict, and at method level the rule is inferior.
\item \textbf{The two populations sit on one raw scale, so the first step is
  well posed.} Pooled, the raw value of the signal ranks god classes better
  than its per-repository standardised score, by a resolved difference on
  both populations, and a cutoff tuned on one population and carried to the
  other lost almost nothing. Both are code written largely by people, so the
  test here is of matching their scale, not of surviving a shift.
\item \textbf{The errors have a shape.} The rule's extra flags are small,
  mostly uncoupled classes, just over half holding a method at PMD's own
  method-level threshold; its misses are large classes whose complexity is
  spread over many moderate methods. Against MLCQ's reviewers no detector
  agrees with the Blob verdicts beyond an MCC of about \agrBlobMaxMcc{}, and
  on Long Method the relative rule has the highest point estimate of the
  detectors tried. A blinded reading of the extra flags by the author and by
  a language model did not agree on what they are, and agreed on one
  question only, whether the class works on other objects' data. Whether a
  concentration of complexity is a god unit is a coupling question, which a
  follow-on study, registered before these numbers existed, takes up.
\end{enumerate}

\section{Background}
\label{sec:related}

Marinescu's detection strategy is a conjunction of metric filters with
thresholds \cite{marinescu2004}; Lanza and Marinescu fixed the GodClass
thresholds from forty-five systems \cite{lanza2006}, DECOR expressed the Blob
as a rule card over size, cohesion and association with data classes
\cite{moha2010}, and PMD implements the Lanza--Marinescu GodClass directly.
Every one of these has a coupling or cohesion clause beside the complexity
clause; this paper uses them as oracles for the complexity clause. That the
constants are hard to justify is an old observation: thresholds have been
derived from benchmarks \cite{alves2010}, fitted distributions
\cite{ferreira2012}, relative positions in a corpus \cite{oliveira2014}, per
language \cite{herbold2011} and per architectural role \cite{aniche2016}.
Each still ships a constant derived from systems other than the one scored, and Baxter et al.\ found the shape of the distributions
those constants cut varying with the application \cite{baxter2006}. The rule
tested here is fitted on each repository's own units at the time of use.

The confound that matters is size: El Emam et al.\ showed most
object-oriented metrics confounded with class size \cite{elemam2001}, and every
per-file sum is nearly collinear with the size clause of every definition. The
study therefore assigns every signal to a family in advance, \emph{size}
(per-file sums and the method count), \emph{peak} (per-file maxima) and
\emph{density} (per-file means) at class level, and the primary may be chosen
from the peak and density families only.

MLCQ is a data set of code smell reviews by professional developers on Java
open-source samples, graded without a given definition and with several
reviewers per sample \cite{mlcq2020}; classifiers trained on it reach a median
MCC of about 0.51 for Blob and 0.81 for Long Method \cite{mlcq2023}, a
reference point rather than a baseline, since the rule here never sees the
labels. Learned detectors are the other route past a hand-set constant:
Arcelli Fontana et al.\ benchmarked sixteen learners on hand-labelled smells
\cite{fontana2016}, and Di Nucci et al.\ showed the figures fell once the
benchmark's balance and overlap were corrected \cite{dinucci2018}. A learned
detector carries its corpus in its weights as a fitted threshold carries it
in a constant; neither is tested here. Schumacher et al.\ found developers agree with one another only
moderately on god classes \cite{schumacher2010}. The reviewers graded samples
without the repository in view, so a human label is an absolute judgement by a
reader's own budget: external, not scale-free, and the one oracle with no
numeric constant inside it.

Flagged sets are scored by the Matthews correlation coefficient
\cite{chicco2020}, signals by the Mann--Whitney AUC, and intervals are
percentile bootstraps that resample repositories \cite{davison1997}. The
pre-registration follows Nosek et al.\ \cite{nosek2018}: the plan was
committed and tagged before the confirmatory sample existed, and everything
outside it is labelled exploratory.

\section{The claim and how it is read}
\label{sec:claim}

The primary rule is the modified z-score,
\begin{equation}
z(x) = \frac{0.6745\,(x - \tilde{x})}{\mathrm{MAD}}, \qquad \text{flag when } z \geq 3.5,
\end{equation}
with $\tilde{x}$ the median and MAD the median absolute deviation of every
unit in the repository at that level, zeros included. When more than half the
units tie at the median the MAD is zero and the rule falls back to the mean
absolute deviation with its normal consistency constant; which scale was used
is recorded, because the rule is scale-free but not shape-free. It is one of
five scale-free rules the plan's Friedman read-out compares by
per-repository MCC against PMD GodClass, and it had the best mean rank on the
main run, \friedRankMadZ{} against \friedRankIqr{} for Tukey's fence and
\friedRankDoubleMad{} for a one-sided MAD, within the critical difference of
\friedRulesCD{} of both, and ahead of the top percentile at
\friedRankPctTopOne{} and the skew-adjusted boxplot at
\friedRankAdjBoxplot{}. The skew-adjusted boxplot flags
\primFlagRateAdjBoxplot{} of Java files against the modified z's
\primFlagRateMadz{}, a fifth as many, and ranks last: the rules that flag
less recover less.

\paragraph{Three readings.} A within-repository AUC is invariant to any
monotone per-repository transform, so it is identical under the relative and
the absolute view: it is evidence about the signal, never about
scale-freeness. The threshold-free reading pools every repository's units into
one ranking, once on the raw value and once on the per-repository modified z,
and compares the two pooled AUCs as a paired difference. The operating-point
reading compares the relative rule at its literature default with the single
raw-scale cutoff that maximised the pooled MCC on the explore half, both on the
confirm half. Every tool definition is itself an absolute threshold, so this
comparison is tilted toward the absolute competitor: a relative win is strong
evidence, a relative loss weak evidence against relative rules in general. A
relative rule also cannot see what an absolute rule uses: in a repository of
uniformly small classes it must still flag something and PMD marks nothing.
Those false positives are reported, and the decision has an arm that sees
them.

\paragraph{The decision rule.} Pre-registered, and read as the weakest of
three paired comparisons of the relative rule at its default minus the tuned
absolute cutoff, each on the same confirm repositories with a 95\% percentile
bootstrap of 2{,}000 resamples over repositories: \emph{macro MCC}, the mean
per-repository MCC difference over the confirm repositories with at least one
positive; \emph{micro MCC}, the pooled MCC difference over every confirm
repository carrying the definition, god-free ones included; and \emph{macro
FPR}, the mean per-repository false-positive-rate difference over every
confirm repository, lower being better. On the MCC arms the relative rule is
\emph{superior} if the lower bound of the difference is above 0,
\emph{non-inferior} if above $-m$, \emph{inferior} if the upper bound is
below $-m$, and \emph{inconclusive} otherwise, with $m = \min(0.10,\ 0.30
\times$ the explore-half macro MCC of the relative rule$)$. On the FPR arm the
rule is non-inferior if the upper bound is below $+0.05$, superior if below 0,
inferior if the lower bound is above $+0.05$. A signal carries the definition
only if the lower bound of the confirm macro AUC is at least 0.70, and the
primary is confirmatory only on at least 20 confirm repositories with three or
more positives and 40 carrying the definition. The power analysis behind these
constants, from a seed-1 pilot of \nPilot{} repositories, is in the registered
plan: a margin of 0.05 was not resolvable at any affordable sample, 0.10 from
about sixty confirm repositories. A margin of 0.10 is about a third of the MCC
levels realised here, so non-inferiority on an MCC arm is a weak bar, and the
point estimates are read beside the verdict for that reason.

\section{Study design}
\label{sec:design}

\paragraph{Sample.} Repositories are drawn by a seeded, stratified sampler, a
separate tool of the author's whose only inputs are GitHub's search index and
criteria fixed before any repository is inspected: ten languages (Java,
Python, JavaScript, TypeScript, Go, C, C++, C\#, PHP and Rust) by three
popularity tiers (above 5{,}000 stars, 1{,}000 to 5{,}000, 200 to 999),
\samplePopulation{} repositories in the frame, shuffled under the
pre-registered seed 2. A candidate is usable on size alone, at least 50
functions in at least 15 source files, decided before any detection outcome.
The author's own ecosystem is excluded. The draw overlaps the seed-1 pilot
in \nPilotOverlap{} repository, \pilotOverlapRepo{}, which is Go, so the
plan's re-read of the primary without the pilot's repositories is the primary
itself. Each language is drawn to 30 usable
repositories and, because every composite class-level definition is Java-only,
a Java supplement is drawn with the same seed to a target of 150, of which
\nJavaSupp{} were usable; with the \nJavaMain{} Java repositories of the main
draw that gives \nJava{} Java repositories among \nUsable{}. A repository is
stratified by the language GitHub indexes it under and counted by the
language found dominant on measurement; \sampleCrossStratumN{} repositories
of the main draw differ on the two, so the counts by measured language sit
near 30 rather than at it. Each repository is pinned to the commit its
default branch was on when drawn. Because every composite definition is
Java-only, the other nine languages enter only the internal Brain Method
comparison and the shape diagnostics of Section~\ref{sec:diag}; every
headline number is a Java number. The \nUsable{} rows have \nOwners{} distinct
GitHub owners, so the bootstrap's assumption of independent repositories is
close to met in the main run; it is not met in the MLCQ run, where
\mlcqTopOwner{} alone holds \mlcqApacheShare{} of the repositories, and MLCQ's
intervals are read with that in mind.

\paragraph{Measurement.} Every function of every source file outside test,
vendored and build directories is measured with lizard \cite{lizard} for
cyclomatic complexity \cite{mcCabe1976}, lines and tokens, and by the
harness's own scanners for cognitive complexity \cite{campbell2018}, nesting,
Halstead volume, maintainability burden, exit points and parameters. Class
level is the file, with each measure aggregated over the file's functions as a
sum, a mean and a maximum, plus the method count; the cyclomatic sum is WMC
\cite{ck1994}. A file is positive under a class definition when any class it
holds meets it; \javaFilesMultiPct{} of Java files hold more than one class.
The lexical estimators track PMD's own values on every Java method at Spearman
\calibCcRho{} for cyclomatic and \calibCognitiveRho{} for cognitive, with AUCs
against PMD's own thresholds of \calibCcAuc{} and above, so a weak result
cannot be blamed on the estimators.

\paragraph{Definitions and rules.} The headline definitions are PMD's GodClass
(tool, Java, class level), PMD's NPath rule at 200 \cite{nejmeh1988} (tool,
Java, method level), an approximate DECOR Blob composed from CK's counts
\cite{cktool} as its LargeClass and LowCohesion clauses alone, the
ControllerClass and association clauses omitted (composed, Java), the study's
own Brain Method (fifty lines, cyclomatic 14 and nesting 5; internal, every
language, read as internal consistency only), and the human labels of MLCQ.
PMD and CK run without an auxiliary classpath; the size of that under-count is
bounded in Section~\ref{sec:deviations}. Every rule is fitted on exactly the
population it scores, the whole per-level universe of the repository with
zeros included. A definition scoped to Java is scored on Java units only.

\paragraph{Split and choices.} Every usable repository is assigned to an
explore or a confirm half from the sample file alone, stratified by language
and tier, and the assignment is recorded in every row before the first
repository is measured. The primary signal is chosen on explore among the peak
and density families by the macro mean per-repository AUC over repositories
with at least three positives; the absolute competitor is the single raw cutoff
maximising the pooled MCC on explore; a tuned relative cut is chosen the same
way as a secondary. All are read on confirm. Each secondary definition gets
the same read-outs and the same paired decision, with no multiplicity
correction and no power to change the primary verdict.

\paragraph{The MLCQ sub-study.} MLCQ 1.1 holds \mlcqSamples{} reviewed
samples for the two smells read here, \mlcqBlobReviewed{} classes for Blob
and \mlcqLmReviewed{} methods for Long Method, each with one or more reviews,
over \mlcqRepos{} Java repositories at 2019 commits \cite{mlcq2020}; \mlcqRunDone{}
could be measured at their labelled commit and \mlcqUsable{} cleared the
substance floor. Its reviewed units are a random draw enriched by re-review of
disputed samples, so the positive rate among them is not a population base
rate. Reviews are converted with a pre-registered rule, a sample is positive
when a strict majority of its reviews grade the smell above \emph{none},
giving \mlcqBlob{} Blob positives among \mlcqBlobReviewed{} reviewed classes
and \mlcqLm{} Long Method positives among \mlcqLmReviewed{} reviewed methods.
The human definitions are scored within the reviewed units only, a class no
reviewer looked at being unknown rather than negative, while every rule is
fitted on the repository's whole Java population; because MLCQ reviewed a
handful of units per repository, the plan decides the human definitions on
the micro arm alone. The same run re-reads the whole main plan against PMD's
GodClass on MLCQ's repositories, a replication on a second population.
Measurement was done by the harness at a committed revision recorded in every
row (\harnessCommit{} for the main run, \mlcqHarnessCommit{} for MLCQ), with
lizard 1.23.0, PMD 7.27.0 and CK 0.7.0 pinned by version and checksum.

\paragraph{Timeline.} The plan was committed and tagged (\texttt{plan-v1}) at
09:20 UTC on 2026-10-18, hours before the confirmatory sample was drawn. The
main run's confirmatory analysis was written once, on \analysedUtc{}; the
sample files and the analysis digest were committed together after the run,
at 01:14 UTC on 2026-10-19; the MLCQ run was analysed on
\mlcqAnalysedUtc{}. That history is the author's own record, with no
third-party timestamp: a reader can check in the published repository that
the tag's commit precedes the commit holding the sample files and the digest,
and nothing more. The follow-on study mentioned in Section~\ref{sec:disc}
registered its plan on the morning of 2026-10-20, hours before the human-label
read-out here was first computed, and its plan records what of this study it
had seen.

\section{Results}

\subsection{Shape, read first}
\label{sec:diag}

At class level the chosen signal's modified z used the MAD scale in
\scaleCogMaxMad{} of repositories. At method level the zero-heavy signals fall
back to the mean absolute deviation far more often: cyclomatic complexity,
which is 1 for more than half of all methods, used the MAD scale in only
\scaleMethodCcMad{} of repositories. The method-level rule is therefore fitted
on populations that tie at the floor, and its flag rate is a property of the
tail's shape as much as of its constant. At class level the default flags
between \flagClassMadMin{} and \flagClassMadMax{} of files in the median
repository of each language: no language sits apart, but a rule that admits a
tenth of the population is not an outlier rule in the usual sense, and the
heavy tail against a small MAD that collapses the method-level rule is present
at class level too, in milder form.

\subsection{The primary: PMD GodClass at class level}
\label{sec:primary}

PMD's GodClass was evaluated on \primCarryEx{} explore and \primCarryCo{}
confirm Java repositories, which carry the definition, PMD having produced no
class verdict on the one remaining confirm repository; \primMinPosCo{} confirm repositories carry three or more
god classes, where a per-repository AUC is read, and \pmdGodFreeShare{} of
Java repositories carry none. Among the peak and density families the
per-file maximum of cognitive complexity, \sig{cognitive\_max}, had the
highest macro AUC on explore, \primExploreAuc{}, ahead of \primExploreSecond{}
at \primExploreSecondAuc{}; its absolute competitor is \sig{cognitive\_max}
$\geq \primCut$. A per-file maximum is a method-level property lifted to class
level: it asks whether the file holds a repository-extreme method, which is
worth holding in mind when the false positives are read
(Section~\ref{sec:fp}).

\paragraph{Read on confirm.} The signal carries the definition: macro AUC
\primAucMacro{}, interval \primAucMacroCI{} over \primAucMacroN{}
repositories, falling to \primAucGiven{} \primAucGivenCI{} within the files
that already pass the WMC $\geq 47$ clause. Much of what the signal separates
is the size clause, and something is left beyond it; the peak correlates with
the sum of lines at Spearman \sizeRhoNlocSum{} per repository. The
threshold-free comparison goes against the relative view: pooled over every
confirm unit the AUC is \primPooledAucRaw{} raw and \primPooledAucStd{}
standardised, a paired difference of \primPooledAucDiff{}
\primPooledAucDiffCI{}, \primPooledAucReading.

\begin{table}[t]
\centering
\caption{The primary read-out on the confirm half: the modified z-score at its
literature default against the absolute cutoff tuned on explore, and the tuned
relative cut as a secondary. Macro is one repository, one vote, over the
\primCarryCo{} confirm repositories carrying the definition, god-free ones at
MCC 0. ``FP in god-free'' is the number of flags in the \primNGodFree{}
confirm repositories where PMD marks nothing.}
\label{tab:confirm-primary}
\small
\setlength{\tabcolsep}{4pt}
\fitwidth{\input{generated/tab_confirm_primary}}
\end{table}

At the operating point (Table~\ref{tab:confirm-primary}) the two rules reach
almost the same macro MCC and the relative rule a lower micro MCC. The
table's macro MCC averages every carrying repository with god-free ones at
0; over the \primMacroMccPosN{} confirm repositories with a god class, the
population of the decision's macro arm and of Table~\ref{tab:transfer}, the
two are \primRelMacroMccPos{} and \primAbsMacroMccPos{}. They get there
differently: the relative rule trades precision for recall, and its
extra flags are false positives, a per-repository false-positive rate of
\primRelFprPct{} against \primAbsFprPct{} and \primRelFpGodFree{} flags in
the \primNGodFree{} god-free repositories against \primAbsFpGodFree{}. The
tuned relative cut, $z \geq \primZCut$, buys recall with more false positives
still; read against the tuned cutoff on the same three arms it is
\transferMainTunedMinusOwnVerdict{}, the verdict the default reaches, so the
constant is not what the rule loses on.

\begin{figure}[t]
\centering
\includegraphics[width=\textwidth]{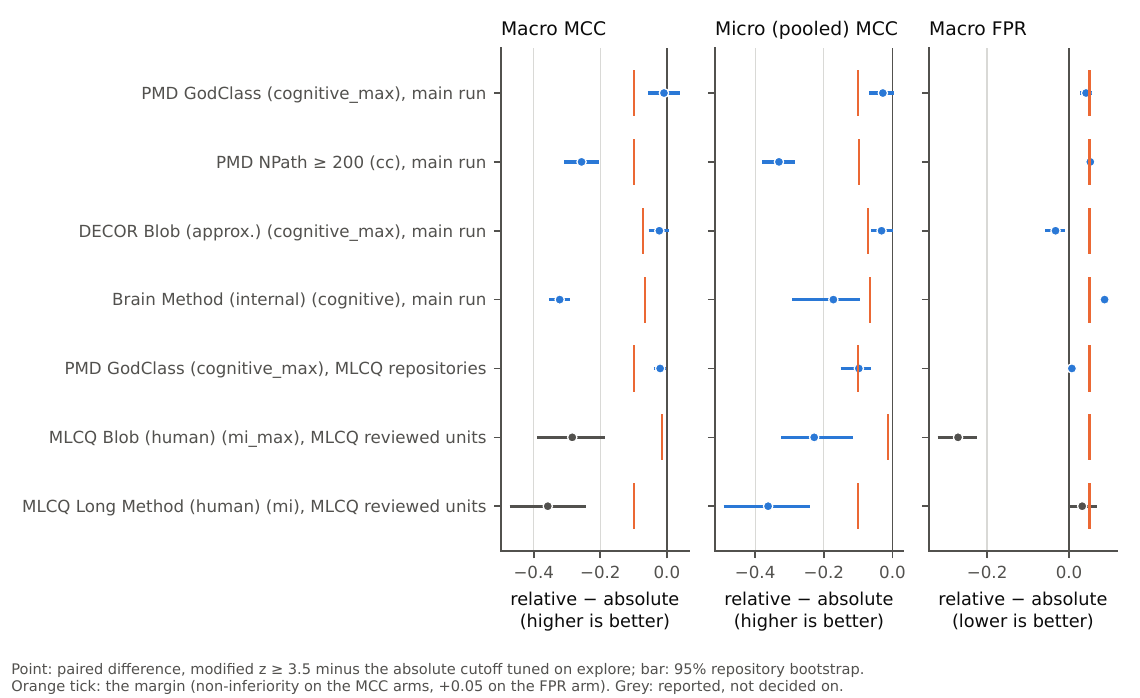}
\caption{The decision arms of Table~\ref{tab:decisions}. On the MCC arms the
margin tick sits at $-m$ and a bar wholly to its right is non-inferior; on the
FPR arm it sits at $+0.05$ and a bar wholly to its left is non-inferior.}
\label{fig:decisions}
\end{figure}

\paragraph{The decision.} Table~\ref{tab:decisions} and
Figure~\ref{fig:decisions} give the three arms. The margin is \primMargin{}.
On the macro MCC arm the paired difference is \primMacroMcc{}
\primMacroMccCI{} over \primMacroMccN{} repositories: \primMacroMccVerdict. On
the micro arm it is \primMicroMcc{} \primMicroMccCI{} over \primMicroMccN{}:
\primMicroMccVerdict. On the false-positive arm the relative rule's per-repository
false-positive rate is \primMacroFpr{} higher, \primMacroFprCI{}: neither below the $+0.05$
margin at its upper bound nor above it at its lower, so \primMacroFprVerdict;
the upper bound misses the margin by less than a hundredth, and the lower
bound, above zero, resolves the extra false positives as real. The verdict,
the weakest arm, is \textbf{\primVerdict}. Every point estimate is on the
absolute side, and non-inferiority on the MCC arms says only that the rule is
not more than 0.10 worse. The effective sample and the AUC floor are met; the
classpath bound is met but out of order (Section~\ref{sec:deviations}), so the
verdict is read as confirmatory with that qualification. The signal ranks
alike in every slice: the macro AUC is \primBrTierHighAuc{},
\primBrTierMidAuc{} and \primBrTierLowAuc{} by popularity tier, and
\primBrKindApplicationAuc{} and \primBrKindLibraryAuc{} in application and
library repositories, and the rule's macro MCC by tier is
\primBrTierHighMcc{}, \primBrTierMidMcc{} and \primBrTierLowMcc{}.

\begin{table}[t]
\centering
\caption{The pre-registered decision for every definition read: the paired
difference, relative rule at its default minus the absolute cutoff tuned on
explore, with its 95\% repository bootstrap, on each arm. The MCC margin is
$\min(0.10, 0.30 \times$ explore level$)$; the FPR margin is $+0.05$. The
verdict is the weakest arm among those the plan decides on; $\dagger$ marks an
arm reported but not decided on (the human definitions decide on the micro arm
alone).}
\label{tab:decisions}
\footnotesize
\setlength{\tabcolsep}{4pt}
\fitwidth{\input{generated/tab_decisions}}
\end{table}

\paragraph{Replication on MLCQ's population.} The same plan, read on the
\mlcqUsable{} MLCQ repositories with their own halves and PMD's GodClass as
the oracle, chooses the same signal at a macro AUC of \mlcqPmdAucMacro{}
\mlcqPmdAucMacroCI{} over \mlcqPmdAucMacroN{} confirm repositories and reaches
the same verdict, \textbf{\mlcqPmdVerdict}, by a different arm: the macro MCC
arm is \mlcqPmdMacroMccVerdict{} (\mlcqPmdMacroMcc{} \mlcqPmdMacroMccCI{},
within the margin though resolved below zero), the FPR arm
\mlcqPmdMacroFprVerdict{} (\mlcqPmdMacroFpr{} \mlcqPmdMacroFprCI{}), and the
micro arm \mlcqPmdMicroMccVerdict{} (\mlcqPmdMicroMcc{} \mlcqPmdMicroMccCI{}).
The tuned cutoff is lower there (\sig{cognitive\_max} $\geq \mlcqPmdCut$
against \primCut{}) and flags about as many files as the relative rule does,
\mlcqPmdAbsFprPct{} against \mlcqPmdRelFprPct{}, while leading it on pooled
MCC by about a tenth; the pooled AUC difference (\mlcqPmdPooledAucDiff{}
\mlcqPmdPooledAucDiffCI{}) says the same. Tuning the relative constant on
MLCQ's own explore half ($z \geq \transferMlcqOwnZCut$) changes nothing:
\transferMlcqTunedMinusOwnVerdict{}.

\paragraph{Secondary definitions.} At method level the rule is
\textbf{\npathVerdict} against PMD's NPath: cyclomatic complexity ranks the
positives at a macro AUC of \npathAucMacro{}, but with more than half of all
methods at cyclomatic 1 the scale collapses to a mean deviation, 3.5 of it is
a low bar, and the rule flags a twentieth of every repository's methods at
precision \npathRelPrecision{} (micro MCC difference \npathMicroMcc{}
\npathMicroMccCI{}). A cut tuned on the relative axis, $z \geq \npathZCut$,
closes much of the gap, so the failure is the literature constant on this
shape of distribution, not the standardisation. The rule is
\textbf{\decorVerdict} against the approximate DECOR Blob, whose tuned cutoff
is a low \sig{cognitive\_max} $\geq \decorCut$ flagging more non-Blob files
than the rule does, and \textbf{\brainVerdict} against the internal Brain
Method for the NPath reason. On MLCQ the NPath and Brain Method verdicts
replicate; the approximate DECOR Blob's does not, where the plan's procedure
chose \mlcqDecorSignal{} on MLCQ's explore half and the verdict is
\mlcqDecorVerdict{} (micro MCC difference \mlcqDecorMicroMcc{}
\mlcqDecorMicroMccCI{}).

\subsection{Exploratory: the cutoff carried across populations}
\label{sec:transfer}

Every comparison above tunes the absolute cutoff on one population's explore
half and reads it on the same population's confirm half, the best case for a
raw-scale constant; the paper's premise, that this tests matching one scale,
holds only if the two populations share one. Table~\ref{tab:transfer} reads
each confirm half four ways: the relative rule at its default, the relative
cut tuned on that population, the cutoff tuned on that population, and the
cutoff tuned on the other population, seven years and a sampling frame away,
carried across unchanged. The read-out was added after review and is
exploratory.

\begin{table}[t]
\centering
\caption{Exploratory: the absolute cutoff on \sig{cognitive\_max} carried
between populations, and the relative cut tuned on each population, against
PMD GodClass on each population's confirm half. Macro MCC is over the
repositories with at least one god class, macro FPR over every repository
carrying the definition.}
\label{tab:transfer}
\small
\setlength{\tabcolsep}{4pt}
\fitwidth{\input{generated/tab_transfer}}
\end{table}

The constant transfers intact. Tuned on the main run the cutoff is
\sig{cognitive\_max} $\geq \transferMainOwnCut$, on MLCQ $\geq
\transferMlcqOwnCut$; read on the main run, MLCQ's cutoff does slightly
better than the main run's own (paired micro MCC difference
\transferMainOtherMinusOwnMicroMcc{} \transferMainOtherMinusOwnMicroMccCI{}),
and read on MLCQ the main run's costs \transferMlcqOtherMinusOwnMicroMcc{}
\transferMlcqOtherMinusOwnMicroMccCI{}. Against the carried constant the
relative rule is behind on the micro arm on both populations, within the
margin on the macro arm (\transferMainRelMinusOtherMacroMcc{}
\transferMainRelMinusOtherMacroMccCI{} on the main run,
\transferMlcqRelMinusOtherMacroMcc{} \transferMlcqRelMinusOtherMacroMccCI{}
on MLCQ), and flags more. What moves is the optimum per repository, not the
pooled one: the cutoff that maximises a single confirm repository's own MCC
runs from \transferMainOptCutMin{} to \transferMainOptCutMax{} on the main
run, quartiles \transferMainOptCutQOne{}, \transferMainOptCutMedian{} and
\transferMainOptCutQThree{}. The relative rule follows that spread, and
following it does not recover PMD's verdicts better than one constant for
everyone, because PMD's definition is itself one constant for everyone.

Read with care, the transfer says less than it seems to. A pooled objective
that cannot tell \transferMainOwnCut{} from \transferMlcqOwnCut{} cannot tell
whether the scale moved by that much either, so the transfer is weak evidence
for a shared scale and no evidence of a shift; it also means the competitor
in the primary was loosely identified, which if anything understates the
rule's deficit. The sharper test is the threshold-free one of
Section~\ref{sec:primary}: pooling every confirm unit, the per-repository
modified z ranks god classes worse than the raw value on both populations,
\primPooledAucDiff{} \primPooledAucDiffCI{} and \mlcqPmdPooledAucDiff{}
\mlcqPmdPooledAucDiffCI{}, both resolved. One raw axis serves every
repository in both populations better than each repository's own. The
comparisons above are therefore a test of matching a single human scale, not
of drift: a constant against an oracle built from the same constant cannot
show the oracle's constant moving, and neither population is agent-written.

\subsection{Human labels}
\label{sec:human}

\paragraph{The planned read-out.} For Blob only \humBlobAllMinPosEx{} explore
repositories have three or more reviewed positives, so the effective-sample
gate (20) is not met, and the plan's procedure chose \humBlobAllSignal{}, the
per-file maximum of the maintainability burden. The choice exposes a flaw in
the plan: the maintainability index is bounded and logarithmic, so its upper
tail is compressed; a per-repository AUC is invariant to that compression,
while a modified z of 3.5 on the compressed axis is out of reach in most
repositories, and the chosen rule cannot flag. The micro MCC difference is
\humBlobAllMicroMcc{} \humBlobAllMicroMccCI{}: \textbf{\humBlobAllVerdict},
\humBlobAllPowered{}. For Long Method the procedure chose
\humLongMethodAllSignal{} on \humLongMethodAllMinPosEx{} repositories, the
same bounded index at method level. There the rule does flag,
\humLongMethodAllRelFlagged{} methods at recall \humLongMethodAllRelRecall{},
but the cutoff tuned on the same axis reaches recall
\humLongMethodAllAbsRecall{}, and the read-out is \humLongMethodAllMicroMcc{}
\humLongMethodAllMicroMccCI{}, \textbf{\humLongMethodAllVerdict} and
\humLongMethodAllPowered{}. On the two
structure signals the plan did not choose, \sig{cognitive} and \sig{cc}, the
default rule reaches a micro MCC within \humCandLongMethodMaxGap{} of a cutoff
tuned on MLCQ's own explore half on both, \humCandLongMethodCcRelMicroMcc{}
against \humCandLongMethodCcAbsMicroMcc{} on \sig{cc}; that is the best of
\humCandLongMethodN{} candidates and upward-biased. No human-label result here is confirmatory.

\paragraph{Tools and rules against the reviewers.} The exploratory comparison
asks how well each fixed definition agrees with the reviewers at all, and
whether the classes the relative rule flags that PMD does not are classes
people call Blob. Nothing here is tuned on MLCQ: the signal, the rule and the
cutoff are the main run's. Repositories on which the tool definition was never
evaluated are dropped, which is why the unit counts in
Table~\ref{tab:agreement} are slightly below the number of reviews that land
on a measured file.

\begin{table}[t]
\centering
\caption{Exploratory: each fixed definition and the main run's two rules as a
predictor of the reviewers' verdict, over the reviewed units of every usable
MLCQ repository, with a 95\% repository bootstrap on the MCC.}
\label{tab:agreement}
\small
\setlength{\tabcolsep}{4pt}
\fitwidth{\input{generated/tab_agreement}}
\end{table}

On Blob every detector is weak and none is separated from the others: PMD's
GodClass reaches an MCC of \agrBlobPmdGodclassMcc{} \agrBlobPmdGodclassMccCI{}
against the \agrBlobPos{} reviewer positives among \agrBlobUnits{} reviewed
classes, the relative rule \agrBlobRelMcc{} \agrBlobRelMccCI{}, and no paired
difference between the relative rule and a fixed definition is resolved. The
reviewers' Blob rate is \agrBlobGrpToolRate{} among the classes PMD flags,
\agrBlobGrpRelNotToolRate{} among the \agrBlobGrpRelNotToolN{} classes the
relative rule flags and PMD does not, and \agrBlobGrpNeitherRelNorToolRate{}
among the classes neither flags: four in five of the rule's extra flags are
not Blobs to the reviewers either. On Long Method the fixed thresholds are
precise and recall little, and the relative rule on \sig{cc} trades precision
for recall and has the highest MCC point estimate of the detectors tried,
\agrLmRelMcc{} \agrLmRelMccCI{}; paired over repositories its lead is
resolved over PMD's NPath, CC $\geq 14$, fifty lines and the tuned cutoff,
and not over cognitive $\geq 15$ (\agrLmRelMinusCognitiveOneFive{}
\agrLmRelMinusCognitiveOneFiveCI{}). A reader should expect this number to
shrink on replication.

\subsection{Exploratory: what the false positives are}
\label{sec:fp}

The rule's extra flags against PMD are the cost the decision turned on. Two
read-outs, both added after review, ask what they are. The first
characterises every one of them by the tool definitions' own component
clauses: over the \fpUnits{} Java class units of the \fpRepos{} confirm
repositories carrying PMD GodClass, Table~\ref{tab:fp-breakdown} splits the
units by the rule's flag and PMD's verdict and gives each group's share
passing PMD's size clause (WMC $\geq 47$), marked by a coupling component
(CK's CBO $\geq 14$ or PMD's CouplingBetweenObjects), marked by a cohesion
component, and holding a method at cognitive complexity 15, PMD's own
method-level threshold, with the group medians.

\begin{table}[t]
\centering
\caption{Exploratory: every Java class unit of the confirm repositories
carrying PMD GodClass, grouped by the relative rule's flag and PMD's verdict,
with the share each group's units meet of PMD's size clause, of the coupling
and cohesion components and of PMD's method-level cognitive threshold, and
the group medians. The last row is the explore-tuned absolute cutoff's own
extra flags.}
\label{tab:fp-breakdown}
\scriptsize
\setlength{\tabcolsep}{3pt}
\fitwidth{\input{generated/tab_fp_breakdown}}
\end{table}

The extra flags are not god classes by size: only \fpFpWmcFourSevenPct{} of
them pass the size clause against \fpTpWmcFourSevenPct{} of the classes PMD
marks, a coupling component marks \fpFpCoupledPct{} against
\fpTpCoupledPct{}, and the cohesion components separate nothing. What does
mark them is PMD's method-level rule: \fpFpCognitiveOneFivePct{} hold a
method at or above cognitive complexity 15, against
\fpNeitherCognitiveOneFivePct{} of unflagged classes. The absolute cutoff's
own extra flags are larger and more often coupled: at a raw cutoff the extra
flags are big classes that missed on data access or cohesion, at a relative
cut they are small classes with one complex method, in the median a file of
about ten methods and, in just over half of cases, a method at PMD's
cognitive threshold in a small class.

The classes the rule misses have the opposite shape. The \fpFnN{} god classes
PMD marks and the rule does not flag have a median of \fpFnMedianNom{}
methods, \fpFnWmcFourSevenPct{} pass the size clause and \fpFnCoupledPct{}
are coupled, but only \fpFnCognitiveOneFivePct{} hold a method at cognitive
15. Their complexity is spread over many moderate methods, and a per-file
maximum cannot see it: the family restriction that keeps the size confound
out of the signal (Section~\ref{sec:related}) also makes it blind to the god
class that is large without being anywhere steep, the cost of reading
concentration as a peak rather than as a share of the repository's total.

The second read-out reads the source. A seeded sample of \fpSampleFp{} of the
rule's extra flags and \fpSampleTp{} of the classes it flags that PMD also
marks, at most \fpSamplePerRepo{} per repository, was read blind to PMD's
verdict, though not to the file's size, under a protocol written before any
file was opened: four yes-or-no questions (whether the most complex method
implements one algorithm over the class's own inputs, whether the class has
one responsibility, whether it works mostly on other objects' data, and
whether its complexity is one dispatch table) and a fixed mapping from the
answers to four categories, leaf, gatherer, table and other. Reading
conventions that settle what the questions leave open were added after the
author had read 13 units and before any model reading; they change no
question or mapping row, and those 13 units were re-read under them. The
author read a seeded subset of \fpReadAuthorFpN{} extra flags and
\fpReadAuthorTpN{} of PMD's god classes before the model read; the model,
Claude Fable 5.1 through the Claude Code command line (version 2.1.307) on
\fpModelReadDate{}, read all \fpSampleN{} in ten independent sessions of
twenty files at default sampling settings, each seeing only the protocol and
its batch. The protocol, every reader's answers and the model's recorded
sessions are in the ancillary data.

\begin{table}[t]
\centering
\caption{Exploratory: the blinded reading of \fpSampleFp{} of the relative
rule's extra flags and \fpSampleTp{} of its flags that PMD also marks, by the
author on a seeded subset and by a language model on every unit, each unit's
category given by the protocol's fixed mapping of the four answers. Leaf: one
algorithm, one responsibility, own data. Gatherer: more than one
responsibility, or mostly other objects' data. Table: one dispatch table.}
\label{tab:fp-readings}
\small
\fitwidth{\input{generated/tab_fp_readings}}
\end{table}

Table~\ref{tab:fp-readings} gives the reading. The two readers do not agree on
what the files are: to the author \fpReadAuthorFpGathererPct{} of the extra
flags are gatherers, to the model \fpReadModelFpLeafPct{} are leaves, and on
the \fpKappaAuthorModelN{} units both read they agree on
\fpKappaAuthorModelAgree{}, Cohen's $\kappa$ of \fpKappaAuthorModel{}. Each
reads the two groups more alike than apart. The disagreement sits in the
responsibility question ($\kappa$ \fpKappaAuthorModelQTwo{}), which the model
answers yes for \fpReadModelFpQTwoYesPct{} of the extra flags and
\fpReadModelTpQTwoYesPct{} of PMD's god classes and the author for
\fpReadAuthorFpQTwoYesPct{} and \fpReadAuthorTpQTwoYesPct{}. The one question
they answer alike is whether the class works mostly on other objects' data
($\kappa$ \fpKappaAuthorModelQThree{}, against \fpKappaAuthorModelQOne{} on
the algorithm question and \fpKappaAuthorModelQFour{} on the dispatch
table): yes for
\fpReadAuthorFpQThreeYesPct{} of the extra flags against
\fpReadAuthorTpQThreeYesPct{} of PMD's god classes to the author and
\fpReadModelFpQThreeYesPct{} against \fpReadModelTpQThreeYesPct{} to the
model, the same direction for both, but neither difference resolved,
Fisher's exact test giving $p = \fpSepAuthorQThreeP{}$ and $p =
\fpSepModelQThreeP{}$. The differences that do resolve are the model's, on
responsibility ($p = \fpSepModelQTwoP{}$) and so on gatherers ($p =
\fpSepModelGathererP{}$), on the question the readers answer least alike;
those are two of sixteen uncorrected tests, eight per reader, and are read as
exploratory. The reading establishes neither that the
extra flags are leaves nor that a reader can tell them from PMD's god classes
by eye; what two readers could judge alike is the coupling question the
follow-on study measures rather than reads.

\section{Deviations from the plan}
\label{sec:deviations}

\paragraph{The classpath validation ran after confirm was read.} PMD and CK
ran without an auxiliary classpath, so GodClass's data-access clause sees only
the sources. The plan bounded that under-count by building at least ten
explore-half Java repositories with their own build tool and re-running PMD
with the built classpath, with a bound of 10\%; it was due before the confirm
half was read and was run after, on explore repositories only. Of \cpTried{}
candidates worked down in seeded order, \cpRepos{} built. \cpFlips{} of
\cpUnion{} verdicts flipped, \cpRate{}, every one a god class gained with the
classpath; \cpMaxFlips{} of them are in one repository (\cpMaxFlipsRepo{},
\cpMaxFlipsRepoRate{}) and over the other nine the rate is \cpRateExclMax{}.
The bound is met. The under-count is one-sided, every flip a god class
gained, but its effect on the paired comparison is not: a gained god class
the rule had flagged turns a false positive into a hit, and one it had not
flagged adds a miss. At \cpRate{} of verdicts, \cpRateExclMax{} outside one
repository, the effect is small in either direction. Because the order was wrong the confirmatory analysis file
records the primary as not confirmatory for want of the validation, and the
re-analysis that applied the bound is a dated exploratory file; the two agree
on every number. A reader who holds to the plan's order may read the primary
as exploratory.

\paragraph{The human read-out is computed by the paper's scripts.} The
harness writes its definition registry before the first repository is
measured, and the MLCQ run's registry holds no human definition, so the
planned human read-out is performed by this paper's scripts with the harness's
statistics module and the plan's constants, on the reviewed units reduced from
the per-unit data. The reduction is checked against the harness's own analysis
file on the main run, where every number it reproduces agrees to nine
decimals. It was first computed on 2026-10-20 and last regenerated on
\humComputedUtc{} with every number unchanged.

\paragraph{Added after review.} Four read-outs were added after drafts were
reviewed, none pre-registered and each labelled exploratory where it appears:
the paired differences in the agreement comparison (computed
\agrComputedUtc{}), the cross-population transfer and the tuned relative cut
on the plan's three arms (\transferComputedUtc{}), the size correlation of
the primary signal, and the characterisation and reading of the false
positives (\fpComputedUtc{}). None changes a pre-registered number.

\section{Discussion}
\label{sec:disc}

\paragraph{Not worse on recovery, at a cost.} Against PMD's GodClass the
scale-free rule did not beat the tuned cutoff on any arm, and every point
estimate of the primary is on the cutoff's side; its one resolved win against
a tool definition, fewer false positives against the approximate DECOR Blob,
is on a secondary. Every tool oracle is a raw-scale constant, so against the
tools these comparisons could at best have shown the rule matching a
constant, and a loss against them is weak evidence, as
Section~\ref{sec:claim} said in advance. The one oracle with no constant
inside it, the reviewers, is read on too few repositories to decide. The
confirmatory verdict therefore answers a narrower question than the title
asks: whether a relative outlier recovers PMD's verdicts as well as a
constant tuned to them. On one population it does, within the margin, and on
the other it falls short on pooled recovery; whether it recovers a
constant-free judgement of a god unit as well as a constant does is left
open.

\paragraph{The steps, and which one this is.} The case for a scale-free
detector is that agent-written code may concentrate complexity differently
from the corpora the constants were fitted on \cite{he2026cursor,
zhang2026llm}; in a lab setting, agents building whole applications have been
seen to centralise logic into single ``manager'' classes \cite{zhu2026smells},
while on files with explicit LLM attribution in real repositories
class-level metrics sit at near parity with human-written code
\cite{mao2026measurement}, so the shift, if there is one, may be small. That
case cannot be tested until populations of agent-written code exist at
scale, and a detector that fails on human-written code would have nothing to
offer them. This study is the first step, on human-written code. The second
is the follow-on coupling study, which asks whether the false positives can
be separated from the god units by what a unit does with other objects'
data. The third, which neither study makes, is to take the refined detector
to agent-written code, by agent adoption as He et al.\ did for complexity
\cite{he2026cursor} or under a fixed task as Zhang, Wen and Tempero did
\cite{zhang2026llm}. Nothing here measures whether the scale moves.

\paragraph{Method level, and the class-level rule's flag rate.} At method
level most methods have the lowest possible value, the robust scale collapses
to a mean deviation, and 3.5 of it is a low bar; yet on the structure signals
the default rule reaches a micro MCC within \humCandLongMethodMaxGap{} of a
tuned cutoff against the reviewers' Long Method verdicts. The constant is the
problem, not the standardisation. The same diagnosis applies, in milder form,
at class level, where the default admits a tenth of files. A scale-free rule
needs a constant fitted to the shape of these distributions; the natural
candidate is the same rule on a logarithmic axis, a modified z of
$\log(1+x)$, which is as scale-free as the rule tested and compresses the tail
that collapses the MAD. It was not run and belongs in a registered follow-on.

\paragraph{What warrants the next study.} The rule's extra flags are small
classes with one steep method, one in six of them a Blob to MLCQ's reviewers,
more than twice the rate among unflagged classes and less than among PMD's
flags; its misses are large, coupled classes whose complexity is nowhere
steep; and reading the source separated neither group from PMD's god classes
to either reader. The reading that fits is that complexity concentration
locates where a codebase's complexity has gathered, and that whether the
place is a god unit is legible neither in the complexity nor, reliably, in a
reader's sense of responsibility. A leaf that holds the complexity of the algorithm it computes,
a parser, a solver, a state machine, is an outlier for its repository and is
where that complexity belongs. Complexity alone cannot tell it from a class
that gathers responsibilities; every definition in this study that does so
adds a coupling clause to do it. The follow-on study asks whether a unit's
coupling divergence, what it hands on to the rest of the system against what
it receives, separates a complex leaf from a god unit among the units
complexity alone flags. Its plan was registered before the human-label
numbers here existed and records what of this study it had seen.

\section{Threats to validity}
\label{sec:threats}

\paragraph{Construct.} Every tool definition is an absolute threshold and
imperfect, so against the tools this measures agreement between detectors on
terms that favour the absolute competitor. The approximate DECOR Blob keeps
two of its four clauses; the Brain Method is internal; cognitive complexity,
Halstead volume, the maintainability index and nesting are lexical estimators
calibrated against PMD. The class-level unit is the file. The human labels are
absolute judgements by readers who did not see the repository and who agreed
with one another only moderately.

\paragraph{Internal.} PMD and CK ran without a classpath; the validation
bounds the effect at \cpRate{} of verdicts on ten repositories and was run out
of order. Test code is excluded by directory conventions that can miss or
over-match. The relative rules are fitted with zeros included, which makes the
zero-heavy method-level signals fall back to a mean-deviation scale. The human
read-out rests on \humBlobAllMinPosEx{} (Blob) and \humLongMethodAllMinPosEx{}
(Long Method) repositories for its signal choice.

\paragraph{External.} GitHub repositories of 200 stars or more, ten languages,
one commit each, with equal allocation across tiers, so pooled numbers are not
population estimates. Every composite definition is Java-only, so the recovery
result is a Java, class-level result. The MLCQ population is MLCQ's choice,
half of it Apache, at 2019 commits, and its intervals overstate its
independence; its reviewed units are enriched by re-review. The sampler is the
author's own tool; its draw is reproducible from the published derivation, and
the published sample files, pinned to commits, are the record.

\paragraph{Conclusion validity.} One primary, one decision rule and one read
of the confirm half were fixed in advance; everything added after review is
labelled exploratory, and the one favourable human-label number is a
best-of-many comparison a reader should expect to shrink. The reading has two
readers, one the author and the other a language model whose answers are not
a human's, who agree on the category at a $\kappa$ of \fpKappaAuthorModel{};
the author read \fpKappaAuthorModelN{} units, the reading was blind to PMD's
verdict and not to size, and a third reader would be needed to say which
reading is the better one. The model is a commercial product whose behaviour
can change between releases, so its archived sessions, not a re-run, are the
record.

\section{Reproducibility and data availability}
\label{sec:repro}

Every number in this paper is produced by a script from the harness's output.
The ancillary data holds the harness's confirmatory analysis files and the
dated exploratory re-analysis, the sample files and frozen explore/confirm
assignments, the per-repository summaries and definition registries, the
converted MLCQ labels, the classpath reports, the paper's own reductions, and
every reader's answers on the false-positive sample with the model reader's
recorded sessions, with the digests as committed. The raw per-unit data
regenerates by running the harness on the published sample files, and the
ancillary README names the public repositories of the harness, the sampler,
the companion study and the follow-on study, with the command behind every
number.

\section{Conclusion}

A unit whose complexity is a robust outlier for its own codebase ranks above
the classes PMD does not call god classes in nine pairs of ten, on two
populations of human-written code that sit on one raw scale, and at its
literature default comes within the pre-registered margin of a cutoff tuned
to that scale on one population and short of it on the other, at about twice
the false positives on the first. The verdict is inconclusive, and the result
that carries forward is the shape of the errors. Concentration finds where
the complexity is. Some of those places are god units and some
are small classes doing the work they were written to do, and a complexity
measure cannot tell the two apart because the difference is not in the
complexity. Two readers of the source could not tell them apart by eye
either, and agreed with each other on one question only, whether the class
works on other objects' data. That is a coupling question, and it warrants
the next study, which measures it; the detector that study refines is the one
to take to agent-written code.

\section*{Competing interests}

The author created and maintains OfficeFloor, a framework whose design rests
on taking data off the control flow, and the construct the follow-on study
tests, coupling divergence, comes from that line of work. This study compares
no frameworks and its sample excludes the author's own ecosystem, but the
reading in Section~\ref{sec:disc}, that the false positives call for a
coupling measure, is the reading the author's interest favours, and a reader
should weigh it as such. The author was one of the two readers of the
false-positive sample; that reading was blind to PMD's verdict, made under a
protocol committed before it, and is reported beside the model's with their
disagreement. The sampler and the companion study are the author's as well.
Against the interest: the plan, the seed and the decision rule were fixed
before the data existed, the verdict that resulted is inconclusive rather than
favourable, every deviation is reported with its effect, and a replication
needs no cooperation from the author.

\section*{Funding}

This work received no external funding and was carried out and paid for by the
author.

\end{document}

%% file: generated/numbers.tex
\newcommand{\agrBlobGrpNeitherRelNorToolRate}{7.1\%}

\newcommand{\agrBlobGrpRelNotToolN}{183}

\newcommand{\agrBlobGrpRelNotToolRate}{16.9\%}

\newcommand{\agrBlobGrpToolRate}{30.9\%}
\newcommand{\agrBlobMaxMcc}{0.24}

\newcommand{\agrBlobPmdGodclassMcc}{+0.242}
\newcommand{\agrBlobPmdGodclassMccCI}{[+0.186, +0.301]}

\newcommand{\agrBlobPos}{204}

\newcommand{\agrBlobRelMcc}{+0.202}
\newcommand{\agrBlobRelMccCI}{[+0.153, +0.254]}

\newcommand{\agrBlobUnits}{1{,}827}

\newcommand{\agrComputedUtc}{2026-10-22 10:04:06 UTC}

\newcommand{\agrLmRelMcc}{+0.582}
\newcommand{\agrLmRelMccCI}{[+0.517, +0.641]}

\newcommand{\agrLmRelMinusCognitiveOneFive}{+0.017}
\newcommand{\agrLmRelMinusCognitiveOneFiveCI}{[$-$0.050, +0.085]}

\newcommand{\analysedUtc}{2026-10-18 22:18:48 UTC}

\newcommand{\brainVerdict}{inferior}

\newcommand{\calibCcAuc}{0.997}
\newcommand{\calibCcRho}{0.95}

\newcommand{\calibCognitiveRho}{0.97}

\newcommand{\cpFlips}{26}

\newcommand{\cpMaxFlips}{23}
\newcommand{\cpMaxFlipsRepo}{Anuken/Mindustry}
\newcommand{\cpMaxFlipsRepoRate}{16.3\%}

\newcommand{\cpRate}{6.7\%}
\newcommand{\cpRateExclMax}{1.2\%}
\newcommand{\cpRepos}{10}

\newcommand{\cpTried}{29}
\newcommand{\cpUnion}{390}

\newcommand{\decorCut}{12}

\newcommand{\decorVerdict}{non-inferior}

\newcommand{\flagClassMadMax}{12.3\%}

\newcommand{\flagClassMadMin}{8.4\%}

\newcommand{\fpComputedUtc}{2026-10-22 22:26:23 UTC}

\newcommand{\fpFnCognitiveOneFivePct}{26.5\%}

\newcommand{\fpFnCoupledPct}{66.8\%}

\newcommand{\fpFnMedianNom}{32}

\newcommand{\fpFnN}{1{,}274}

\newcommand{\fpFnWmcFourSevenPct}{91.9\%}

\newcommand{\fpFpCognitiveOneFivePct}{56.3\%}

\newcommand{\fpFpCoupledPct}{38.5\%}

\newcommand{\fpFpWmcFourSevenPct}{24.1\%}

\newcommand{\fpKappaAuthorModel}{0.20}
\newcommand{\fpKappaAuthorModelAgree}{21}
\newcommand{\fpKappaAuthorModelN}{40}
\newcommand{\fpKappaAuthorModelQFour}{0.22}

\newcommand{\fpKappaAuthorModelQOne}{0.23}

\newcommand{\fpKappaAuthorModelQThree}{0.61}

\newcommand{\fpKappaAuthorModelQTwo}{0.07}

\newcommand{\fpModelReadDate}{2026-10-22}

\newcommand{\fpNeitherCognitiveOneFivePct}{2.4\%}

\newcommand{\fpReadAuthorFpGathererPct}{75\%}

\newcommand{\fpReadAuthorFpN}{20}

\newcommand{\fpReadAuthorFpQThreeYesPct}{40\%}

\newcommand{\fpReadAuthorFpQTwoYesPct}{30\%}

\newcommand{\fpReadAuthorTpN}{20}

\newcommand{\fpReadAuthorTpQThreeYesPct}{70\%}

\newcommand{\fpReadAuthorTpQTwoYesPct}{5\%}

\newcommand{\fpReadModelFpLeafPct}{66\%}

\newcommand{\fpReadModelFpQThreeYesPct}{16\%}

\newcommand{\fpReadModelFpQTwoYesPct}{96\%}

\newcommand{\fpReadModelTpQThreeYesPct}{22\%}

\newcommand{\fpReadModelTpQTwoYesPct}{82\%}

\newcommand{\fpRepos}{84}
\newcommand{\fpSampleFp}{100}
\newcommand{\fpSampleN}{200}
\newcommand{\fpSamplePerRepo}{3}
\newcommand{\fpSampleTp}{100}

\newcommand{\fpSepAuthorQThreeP}{0.111}

\newcommand{\fpSepModelGathererP}{0.045}

\newcommand{\fpSepModelQThreeP}{0.368}
\newcommand{\fpSepModelQTwoP}{0.003}

\newcommand{\fpTpCoupledPct}{74.3\%}

\newcommand{\fpTpWmcFourSevenPct}{96.8\%}

\newcommand{\fpUnits}{60{,}363}
\newcommand{\friedRankAdjBoxplot}{3.85}
\newcommand{\friedRankDoubleMad}{2.67}
\newcommand{\friedRankIqr}{2.50}
\newcommand{\friedRankMadZ}{2.42}
\newcommand{\friedRankPctTopOne}{3.56}
\newcommand{\friedRulesCD}{0.53}

\newcommand{\harnessCommit}{f9b26ba}

\newcommand{\humBlobAllMicroMcc}{$-$0.228}
\newcommand{\humBlobAllMicroMccCI}{[$-$0.324, $-$0.115]}

\newcommand{\humBlobAllMinPosEx}{9}

\newcommand{\humBlobAllPowered}{underpowered}

\newcommand{\humBlobAllSignal}{\texttt{mi\_max}}
\newcommand{\humBlobAllVerdict}{inferior}

\newcommand{\humCandLongMethodCcAbsMicroMcc}{0.674}
\newcommand{\humCandLongMethodCcRelMicroMcc}{0.617}

\newcommand{\humCandLongMethodMaxGap}{0.08}
\newcommand{\humCandLongMethodN}{7}

\newcommand{\humComputedUtc}{2026-10-22 10:03:52 UTC}

\newcommand{\humLongMethodAllAbsRecall}{0.91}

\newcommand{\humLongMethodAllMicroMcc}{$-$0.362}
\newcommand{\humLongMethodAllMicroMccCI}{[$-$0.492, $-$0.242]}

\newcommand{\humLongMethodAllMinPosEx}{13}

\newcommand{\humLongMethodAllPowered}{underpowered}
\newcommand{\humLongMethodAllRelFlagged}{104}

\newcommand{\humLongMethodAllRelRecall}{0.47}

\newcommand{\humLongMethodAllSignal}{\texttt{mi}}
\newcommand{\humLongMethodAllVerdict}{inferior}

\newcommand{\javaFilesMultiPct}{23\%}
\newcommand{\mlcqAnalysedUtc}{2026-10-19 22:20:30 UTC}
\newcommand{\mlcqApacheShare}{50.2\%}
\newcommand{\mlcqBlob}{254}
\newcommand{\mlcqBlobReviewed}{2{,}334}

\newcommand{\mlcqDecorMicroMcc}{$-$0.082}
\newcommand{\mlcqDecorMicroMccCI}{[$-$0.115, $-$0.052]}
\newcommand{\mlcqDecorSignal}{\texttt{nloc\_max}}
\newcommand{\mlcqDecorVerdict}{inconclusive}

\newcommand{\mlcqHarnessCommit}{7b52774}
\newcommand{\mlcqLm}{276}
\newcommand{\mlcqLmReviewed}{2{,}430}

\newcommand{\mlcqPmdAbsFprPct}{8.0\%}

\newcommand{\mlcqPmdAucMacro}{0.908}
\newcommand{\mlcqPmdAucMacroCI}{[0.900, 0.916]}
\newcommand{\mlcqPmdAucMacroN}{198}

\newcommand{\mlcqPmdCut}{17}
\newcommand{\mlcqPmdMacroFpr}{+0.007}
\newcommand{\mlcqPmdMacroFprCI}{[$-$0.005, +0.018]}
\newcommand{\mlcqPmdMacroFprVerdict}{non-inferior}
\newcommand{\mlcqPmdMacroMcc}{$-$0.020}
\newcommand{\mlcqPmdMacroMccCI}{[$-$0.039, $-$0.003]}

\newcommand{\mlcqPmdMacroMccVerdict}{non-inferior}

\newcommand{\mlcqPmdMicroMcc}{$-$0.098}
\newcommand{\mlcqPmdMicroMccCI}{[$-$0.151, $-$0.063]}
\newcommand{\mlcqPmdMicroMccVerdict}{inconclusive}

\newcommand{\mlcqPmdPooledAucDiff}{$-$0.045}
\newcommand{\mlcqPmdPooledAucDiffCI}{[$-$0.080, $-$0.024]}

\newcommand{\mlcqPmdRelFprPct}{8.7\%}

\newcommand{\mlcqPmdVerdict}{inconclusive}

\newcommand{\mlcqRepos}{522}
\newcommand{\mlcqRunDone}{499}

\newcommand{\mlcqSamples}{4{,}764}
\newcommand{\mlcqTopOwner}{apache}

\newcommand{\mlcqUsable}{492}
\newcommand{\nJava}{171}
\newcommand{\nJavaMain}{30}
\newcommand{\nJavaSupp}{141}

\newcommand{\nOwners}{430}
\newcommand{\nPilot}{522}
\newcommand{\nPilotOverlap}{1}

\newcommand{\nUsable}{445}

\newcommand{\npathAucMacro}{0.996}

\newcommand{\npathMicroMcc}{$-$0.331}
\newcommand{\npathMicroMccCI}{[$-$0.380, $-$0.284]}

\newcommand{\npathRelPrecision}{0.07}

\newcommand{\npathVerdict}{inferior}
\newcommand{\npathZCut}{8.75}
\newcommand{\pilotOverlapRepo}{runkids/skillshare}
\newcommand{\pmdGodFreeShare}{22\%}

\newcommand{\primAbsFpGodFree}{19}

\newcommand{\primAbsFprPct}{3.8\%}

\newcommand{\primAbsMacroMccPos}{0.380}

\newcommand{\primAucGiven}{0.624}
\newcommand{\primAucGivenCI}{[0.579, 0.672]}
\newcommand{\primAucMacro}{0.916}
\newcommand{\primAucMacroCI}{[0.903, 0.928]}
\newcommand{\primAucMacroN}{51}
\newcommand{\primBrKindApplicationAuc}{0.920}

\newcommand{\primBrKindLibraryAuc}{0.904}

\newcommand{\primBrTierHighAuc}{0.911}
\newcommand{\primBrTierHighMcc}{0.297}

\newcommand{\primBrTierLowAuc}{0.916}
\newcommand{\primBrTierLowMcc}{0.253}

\newcommand{\primBrTierMidAuc}{0.920}
\newcommand{\primBrTierMidMcc}{0.278}

\newcommand{\primCarryCo}{84}
\newcommand{\primCarryEx}{86}
\newcommand{\primCut}{22}
\newcommand{\primExploreAuc}{0.916}

\newcommand{\primExploreSecond}{\texttt{nloc\_max}}
\newcommand{\primExploreSecondAuc}{0.893}
\newcommand{\primFlagRateAdjBoxplot}{2.1\%}

\newcommand{\primFlagRateMadz}{10.6\%}

\newcommand{\primMacroFpr}{+0.042}
\newcommand{\primMacroFprCI}{[+0.027, +0.056]}

\newcommand{\primMacroFprVerdict}{inconclusive}
\newcommand{\primMacroMcc}{$-$0.009}
\newcommand{\primMacroMccCI}{[$-$0.057, +0.040]}
\newcommand{\primMacroMccN}{63}
\newcommand{\primMacroMccPosN}{63}
\newcommand{\primMacroMccVerdict}{non-inferior}
\newcommand{\primMargin}{0.10}
\newcommand{\primMicroMcc}{$-$0.028}
\newcommand{\primMicroMccCI}{[$-$0.068, +0.005]}
\newcommand{\primMicroMccN}{84}
\newcommand{\primMicroMccVerdict}{non-inferior}
\newcommand{\primMinPosCo}{51}

\newcommand{\primNGodFree}{21}

\newcommand{\primPooledAucDiff}{$-$0.020}
\newcommand{\primPooledAucDiffCI}{[$-$0.033, $-$0.008]}
\newcommand{\primPooledAucRaw}{0.923}
\newcommand{\primPooledAucReading}{absolute ranking better}
\newcommand{\primPooledAucStd}{0.903}

\newcommand{\primRelFpGodFree}{139}

\newcommand{\primRelFprPct}{7.9\%}

\newcommand{\primRelMacroMccPos}{0.371}

\newcommand{\primVerdict}{inconclusive}
\newcommand{\primZCut}{2.75}

\newcommand{\sampleCrossStratumN}{30}

\newcommand{\samplePopulation}{170{,}596}

\newcommand{\scaleCogMaxMad}{89\%}

\newcommand{\scaleMethodCcMad}{34\%}

\newcommand{\sizeRhoNlocSum}{0.70}

\newcommand{\transferComputedUtc}{2026-10-22 16:59:08 UTC}

\newcommand{\transferMainOptCutMax}{85}
\newcommand{\transferMainOptCutMedian}{15}
\newcommand{\transferMainOptCutMin}{7}

\newcommand{\transferMainOptCutQOne}{12}
\newcommand{\transferMainOptCutQThree}{26}

\newcommand{\transferMainOtherMinusOwnMicroMcc}{+0.015}
\newcommand{\transferMainOtherMinusOwnMicroMccCI}{[$-$0.003, +0.034]}

\newcommand{\transferMainOwnCut}{22}

\newcommand{\transferMainRelMinusOtherMacroMcc}{$-$0.030}
\newcommand{\transferMainRelMinusOtherMacroMccCI}{[$-$0.062, +0.002]}

\newcommand{\transferMainTunedMinusOwnVerdict}{inconclusive}

\newcommand{\transferMlcqOtherMinusOwnMicroMcc}{$-$0.007}
\newcommand{\transferMlcqOtherMinusOwnMicroMccCI}{[$-$0.020, +0.006]}

\newcommand{\transferMlcqOwnCut}{17}
\newcommand{\transferMlcqOwnZCut}{2.87}

\newcommand{\transferMlcqRelMinusOtherMacroMcc}{$-$0.006}
\newcommand{\transferMlcqRelMinusOtherMacroMccCI}{[$-$0.025, +0.011]}

\newcommand{\transferMlcqTunedMinusOwnVerdict}{inconclusive}

%% file: generated/tab_confirm_primary.tex
% generated by paper/scripts/tables.py; do not edit
\begin{tabular}{lrrrrrr}
\toprule
Rule on \texttt{cognitive\_max} & Macro MCC, god-free at 0 (95\% CI) & Micro MCC & Precision & Recall & Macro FPR & FP in god-free repos \\
\midrule
default: modified $z \geq 3.5$ & 0.278 [0.230, 0.329] & 0.378 & 0.30 & 0.59 & 0.079 & 139/21 \\
tuned absolute cutoff $\geq 22$ & 0.285 [0.229, 0.341] & 0.406 & 0.44 & 0.44 & 0.038 & 19/21 \\
tuned modified $z \geq 2.75$ (secondary) & 0.297 [0.248, 0.346] & 0.380 & 0.27 & 0.67 & 0.102 & 203/21 \\
\bottomrule
\end{tabular}

%% file: generated/tab_decisions.tex
% generated by paper/scripts/tables.py; do not edit
\begin{tabular}{lcccrl}
\toprule
Definition (signal) & Macro MCC & Micro MCC & Macro FPR & Margin & Verdict \\
\midrule
PMD GodClass & $-$0.009 & $-$0.028 & +0.042 & 0.100 & \textbf{inconclusive} \\
{\scriptsize \texttt{cognitive\_max}, main run} & {\scriptsize [$-$0.057, +0.040] non-inferior} & {\scriptsize [$-$0.068, +0.005] non-inferior} & {\scriptsize [+0.027, +0.056] inconclusive} & & \\[2pt]
PMD NPath $\geq 200$ & $-$0.256 & $-$0.331 & +0.052 & 0.098 & \textbf{inferior} \\
{\scriptsize \texttt{cc}, main run} & {\scriptsize [$-$0.308, $-$0.204] inferior} & {\scriptsize [$-$0.380, $-$0.284] inferior} & {\scriptsize [+0.047, +0.057] inconclusive} & & \\[2pt]
DECOR Blob (approx.) & $-$0.023 & $-$0.032 & $-$0.033 & 0.072 & \textbf{non-inferior} \\
{\scriptsize \texttt{cognitive\_max}, main run} & {\scriptsize [$-$0.055, +0.006] non-inferior} & {\scriptsize [$-$0.063, +0.001] non-inferior} & {\scriptsize [$-$0.058, $-$0.011] superior} & & \\[2pt]
Brain Method (internal) & $-$0.322 & $-$0.172 & +0.087 & 0.066 & \textbf{inferior} \\
{\scriptsize \texttt{cognitive}, main run} & {\scriptsize [$-$0.355, $-$0.290] inferior} & {\scriptsize [$-$0.292, $-$0.095] inferior} & {\scriptsize [+0.078, +0.096] inferior} & & \\[2pt]
PMD GodClass & $-$0.020 & $-$0.098 & +0.007 & 0.100 & \textbf{inconclusive} \\
{\scriptsize \texttt{cognitive\_max}, MLCQ repositories} & {\scriptsize [$-$0.039, $-$0.003] non-inferior} & {\scriptsize [$-$0.151, $-$0.063] inconclusive} & {\scriptsize [$-$0.005, +0.018] non-inferior} & & \\[2pt]
MLCQ Blob (human) & $-$0.284 & $-$0.228 & $-$0.271 & 0.014 & \textbf{inferior} \\
{\scriptsize \texttt{mi\_max}, MLCQ reviewed units} & {\scriptsize [$-$0.389, $-$0.186] inferior$^{\dagger}$} & {\scriptsize [$-$0.324, $-$0.115] inferior} & {\scriptsize [$-$0.321, $-$0.225] superior$^{\dagger}$} & & \\[2pt]
MLCQ Long Method (human) & $-$0.358 & $-$0.362 & +0.032 & 0.100 & \textbf{inferior} \\
{\scriptsize \texttt{mi}, MLCQ reviewed units} & {\scriptsize [$-$0.473, $-$0.242] inferior$^{\dagger}$} & {\scriptsize [$-$0.492, $-$0.242] inferior} & {\scriptsize [+0.000, +0.069] inconclusive$^{\dagger}$} & & \\[2pt]
\bottomrule
\end{tabular}

%% file: generated/tab_transfer.tex
% generated by paper/scripts/tables.py; do not edit
\begin{tabular}{llrrrrrr}
\toprule
Read on & Rule & Cut & Micro MCC & Macro MCC, $\geq 1$ positive & Macro FPR & Precision & Recall \\
\midrule
main run, confirm half & relative, modified $z \geq 3.5$ & -- & 0.378 & 0.371 & 0.079 & 0.30 & 0.59 \\
 & relative, $z$ tuned on this population & $z \geq 2.75$ & 0.380 & 0.396 & 0.102 & 0.27 & 0.67 \\
 & absolute, tuned on this population & 22 & 0.406 & 0.380 & 0.038 & 0.44 & 0.44 \\
 & absolute, tuned on the MLCQ & 17 & 0.421 & 0.401 & 0.063 & 0.37 & 0.56 \\
\addlinespace
MLCQ, confirm half & relative, modified $z \geq 3.5$ & -- & 0.385 & 0.370 & 0.087 & 0.35 & 0.56 \\
 & relative, $z$ tuned on this population & $z \geq 2.87$ & 0.387 & 0.371 & 0.110 & 0.32 & 0.63 \\
 & absolute, tuned on this population & 17 & 0.483 & 0.390 & 0.080 & 0.41 & 0.69 \\
 & absolute, tuned on the main run & 22 & 0.476 & 0.376 & 0.053 & 0.47 & 0.57 \\
\addlinespace
\bottomrule
\end{tabular}

%% file: generated/tab_agreement.tex
% generated by paper/scripts/tables.py; do not edit
\begin{tabular}{lrrrrl}
\toprule
Predictor & Flagged & Precision & Recall & MCC & 95\% CI \\
\midrule
\multicolumn{6}{l}{\emph{Blob}: 390 repositories, 1{,}827 reviewed units, 204 reviewer positives (base rate 11.2\%)} \\
\quad PMD GodClass & 236 & 0.31 & 0.36 & +0.242 & [+0.186, +0.301] \\
\quad DECOR Blob (approx.) & 335 & 0.24 & 0.40 & +0.196 & [+0.142, +0.256] \\
\quad WMC $\geq 47$ & 398 & 0.25 & 0.48 & +0.226 & [+0.171, +0.283] \\
\quad relative: modified $z \geq 3.5$ on \texttt{cognitive\_max} & 326 & 0.25 & 0.40 & +0.202 & [+0.153, +0.254] \\
\quad absolute: \texttt{cognitive\_max} $\geq 22$ & 242 & 0.29 & 0.34 & +0.220 & [+0.162, +0.284] \\
\addlinespace
\multicolumn{6}{l}{\emph{Long Method}: 392 repositories, 1{,}988 reviewed units, 220 reviewer positives (base rate 11.1\%)} \\
\quad PMD NPath $\geq 200$ & 28 & 0.93 & 0.12 & +0.312 & [+0.244, +0.371] \\
\quad CC $\geq 14$ & 47 & 0.96 & 0.20 & +0.420 & [+0.353, +0.485] \\
\quad cognitive $\geq 15$ & 102 & 0.87 & 0.40 & +0.565 & [+0.497, +0.629] \\
\quad NLOC $\geq 50$ & 68 & 0.91 & 0.28 & +0.481 & [+0.412, +0.542] \\
\quad Brain Method (internal) & 4 & 1.00 & 0.02 & +0.127 & [+0.060, +0.189] \\
\quad relative: modified $z \geq 3.5$ on \texttt{cc} & 205 & 0.65 & 0.60 & +0.582 & [+0.517, +0.641] \\
\quad absolute: \texttt{cc} $\geq 15$ & 45 & 0.98 & 0.20 & +0.421 & [+0.354, +0.485] \\
\addlinespace
\bottomrule
\end{tabular}

%% file: generated/tab_fp_breakdown.tex
% generated by paper/scripts/tables.py; do not edit
\begin{tabular}{lrrrrrrrrr}
\toprule
Group & Units & Repos. & WMC $\geq 47$ & Coupled & Low cohesion & Cog.\ $\geq 15$ & Median \texttt{cognitive\_max} & Median lines & Median methods \\
\midrule
flagged by the rule and marked by PMD & 1{,}841 & 56 & 96.8\% & 74.3\% & 86.5\% & 87.9\% & 29 & 415 & 32 \\
flagged by the rule, not marked by PMD & 4{,}288 & 78 & 24.1\% & 38.5\% & 84.1\% & 56.3\% & 17 & 113 & 9 \\
marked by PMD, not flagged by the rule & 1{,}274 & 50 & 91.9\% & 66.8\% & 91.0\% & 26.5\% & 11 & 272 & 32 \\
neither & 52{,}960 & 84 & 2.5\% & 8.2\% & 78.4\% & 2.4\% & 1 & 27 & 5 \\
flagged by the absolute cutoff, not marked by PMD & 1{,}753 & 64 & 40.6\% & 45.1\% & 83.1\% & 100.0\% & 30 & 165 & 10 \\
\bottomrule
\end{tabular}

%% file: generated/tab_fp_readings.tex
% generated by paper/scripts/tables.py; do not edit
\begin{tabular}{llrrrrr}
\toprule
Reader & Group & $n$ & Leaf & Gatherer & Table & Other \\
\midrule
author & flagged by the rule, not marked by PMD & 20 & 5 (25\%) & 15 (75\%) & 0 (0\%) & 0 (0\%) \\
author & flagged by the rule and marked by PMD & 20 & 1 (5\%) & 19 (95\%) & 0 (0\%) & 0 (0\%) \\
model & flagged by the rule, not marked by PMD & 100 & 66 (66\%) & 17 (17\%) & 5 (5\%) & 12 (12\%) \\
model & flagged by the rule and marked by PMD & 100 & 58 (58\%) & 30 (30\%) & 2 (2\%) & 10 (10\%) \\
\multicolumn{7}{l}{agreement author with model: 21 of 40 units, Cohen's $\kappa = 0.20$} \\
\bottomrule
\end{tabular}